\documentclass[a4paper,fleqn]{cas-dc}

\usepackage[numbers]{natbib}

\def\tsc#1{\csdef{#1}{\textsc{\lowercase{#1}}\xspace}}
\tsc{WGM}
\tsc{QE}
\tsc{EP}
\tsc{PMS}
\tsc{BEC}
\tsc{DE}

\begin{document}
\shorttitle{Calibration system for the Large Sized Telescope camera}
\shortauthors{M. Iori et~al.}

\title [mode = title]{Test results of the optical calibration system for the Large Sized Telescope camera}                      



\author[1]{M. Iori}
\cormark[1]


\affiliation[1]{organization={Istituto Nazionale di Fisica Nucleare Sezione di Roma La Sapienza},
                addressline={P.zza A. Moro 5}, 
                city={Rome},
                postcode={00100}, 
                state={Italy}}

\author[1]{F. Ferrarotto}
\author[1]{L. Recchia}

\author[1]{A .Girardi}
\author[1]{R. Lunadei}

\cortext[cor1]{Corresponding author: maurizio.iori@roma1.infn.it}


\begin{abstract}
In 2018 the Large Sized Telescope (LST-1) prototype, designed to be the lowest energy detector for the Cherenkov Telescope Array  Observatory, was inaugurated at the Observatorio de Roque de Los Muchachos in La Palma, Canary Island and  today three more are under construction, LST2-4. The LST camera, with 1855 photomultipliers (PMTs), requires  precise and  regular calibration. 
The camera calibration system (hereafter CaliBox), installed at the center of the telescope mirror dish, is equipped with a Q-switching 355 nm UV laser corresponding to the wavelength at which the maximum camera PMT quantum efficiency is achieved, a set of filters to guarantee a large dynamic range of photons on each camera pixel, and a Ulbricht sphere to spread uniformly the laser light over the camera plane 28 m away.  The system is managed by an ODROID-C1+ single board computer that  communicates through an Open Platform Communication Unified Architecture (OPCUA) protocol to the camera. The CaliBox is designed to fulfill  the requirements needed for the calibration of the camera including  the monitor of the photon flux to guarantee the quality of the CaliBox system of laser stability, uniform illumination and intensity range.  In this paper, we present  in detail the optical system, the monitor of the photon flux, the relevant electronic to monitor the device.  The performance  of the device, the photon flux monitoring, the evaluation of the photon flux  sent to the camera obtained during  tests  performed in Laboratory are shown.
\end{abstract}

\begin{keywords}
Cherenkov telescope \sep calibration device \sep gamma astronomy
\end{keywords}

\maketitle

\section{Introduction}

 The Cherenkov Telescope Array (CTA)  is the next generation ground-based observatory for $\gamma$-ray astronomy at very high energies up to TeV. With 80 telescopes located in the northern and southern hemisphere,  CTA will be the  world's largest and the most sensitive high energy $\gamma$ ray observatory \cite{CTAph}. 
In order to collect the Cherenkov light generated by the  $\gamma$ rays in the widest energy range, 
 the CTA  plans to build telescopes of different sizes in the Northern site, located at the Observatorio del Roque de los Muchachos, ORM, in La Palma, Canary Island and in the Southern site, Paranal Chile.  At ORM,  the first large-sized telescope, LST-1,  is currently under commissioning  with a parabolic dish of 23 m of diameter to detect the lowest energy $\gamma$ ray events above 20 GeV.
In order to deliver stable and reliable performance during  its livetime, the camera response must be calibrated continuiously.
The absolute charge calibration of the camera PMTs is performed using the F-factor method, which relies on a flat-field events produced by the uniform illumination of the camera  \cite{Factor}. For this purpose, periodic UV light calibration of all PMTs is essential to ensure a flat-field responce. To achive uniform light distribution at the camera plane and cover the full pixel dynamic range from 1 to 3000 p.e., as required by the camera readout electronics, we developed a calibration system called CaliBox. Installed in the fall of 2018 at the center of the LST-1 mirror dish, this system is used to calibrate the PMTs before and during the event acquisition  \cite{doc1} .
This paper aims to describe the calibration system and show the results obtained in the Laboratory. The outline is the following: in Sec. 2 the mechanical frame, electronics and optical system of CaliBox  are described. Sec. 3 discusses the  estimation of the photon density variability, Secs. 4, 5 and 6 present the test results of the flat fielding uniformity measurements, the internal calibration results and the CaliBox thermal conductivity respectively.  

\section{The CaliBox device}
The concept with which the CaliBox design is based upon keeping the optical components separate from the control electronics to ensure  the photon flux required is not altered by the variation of parameters in temperature and relative humidity,  \cite{doc2}.

The description of the CaliBox device is subdivided in four parts: main frame, optical system, monitor of photon flux sent to the camera and the electronic system.

\subsection{Main frame}
\begin{figure}
\centering
\includegraphics[width=.9\columnwidth]{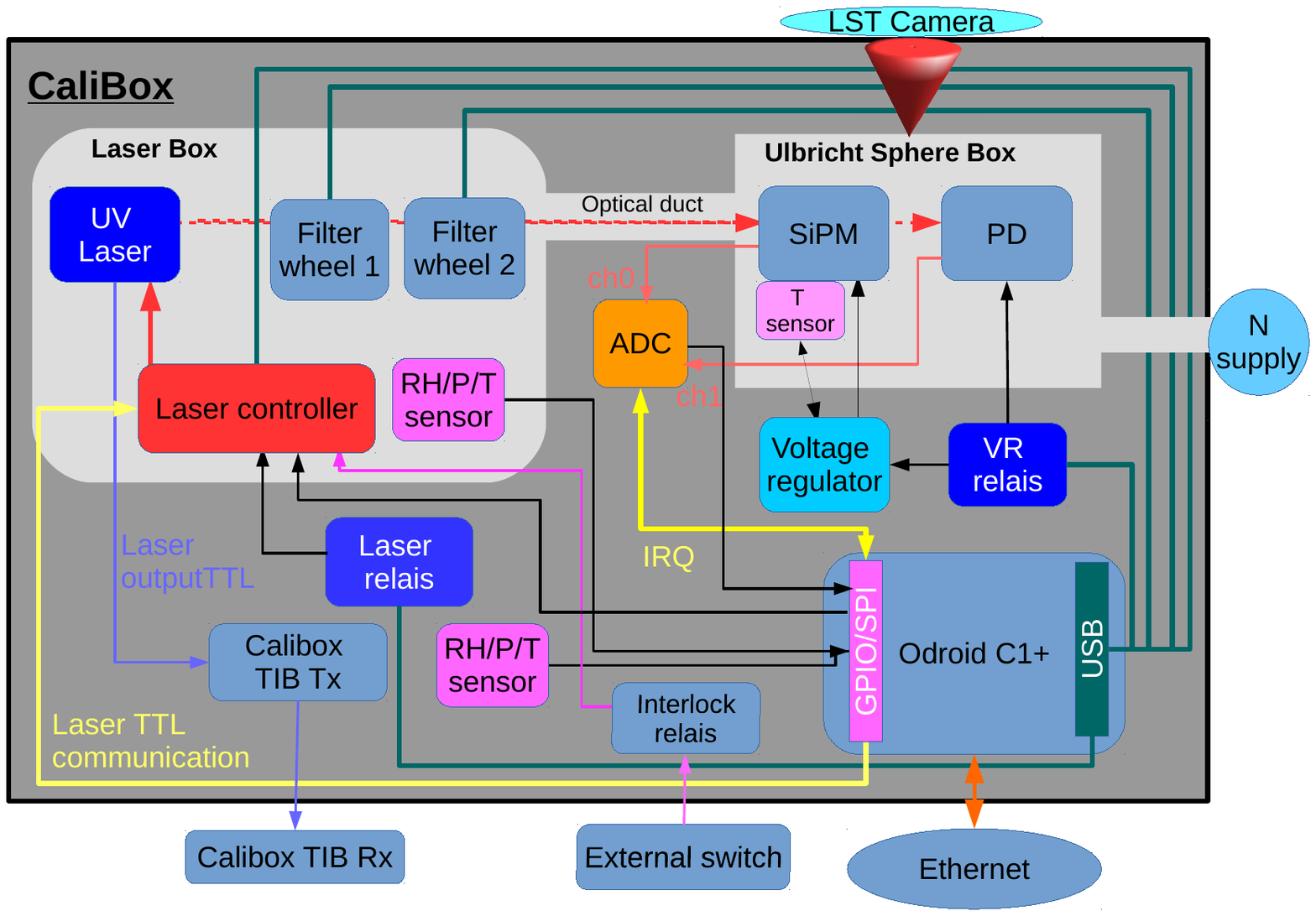}
\caption{CaliBox block diagram. The light gray squares refers to the optical components: laser and Ulbricht sphere. On darked grey background is drawn the ODROID C1+ and the electronic subsystem components. The relative humidity, temperature and the pressure sensors RH/P/T are shown. The ODROID C1+ manages the subsystem by USB (green) and general purpose Input/Output, GPIO (lilac). The blue line shows the TTL laser trigger  signal to be sent to the Camera TIB after is transformed into optical. The ODROID C1+ generates  the interrupt request (IRQ)  to get the ADC data. The connection block diagram is discussed in Sec.2.4}  
\label{fig:optical_path0}
\end{figure}

The main frame of the CaliBox is composed of two aluminum boxes fixed on an aluminum plate and connected together by a vacuum tight connector as shown in the schematic block of  Fig.\ref{fig:optical_path0}.  One box contains the laser with its controller \cite{laser} and two filter wheels \cite{wheel} used to vary the number of photons sent to the camera. The other box contains a 1-inch diameter 3-hole Ulbricht sphere \cite{sphere}, a beam splitter and two different sensors to monitor the photon flux sent to the camera  as described in Sec. 2.3. These two boxes are  hermetically sealed by O-rings and filled with Nitrogen at atmospheric pressure to control the internal relative humidity (RH) and to avoid  oxidation of the internal electronic components and water vapor condensation inside the device. The choice of aluminum is due to its lightness and good thermal conductivity  to dissipate the heat generated by the internal electronic components and the laser.

\begin{figure}
\centering
\includegraphics[width=.8\columnwidth]{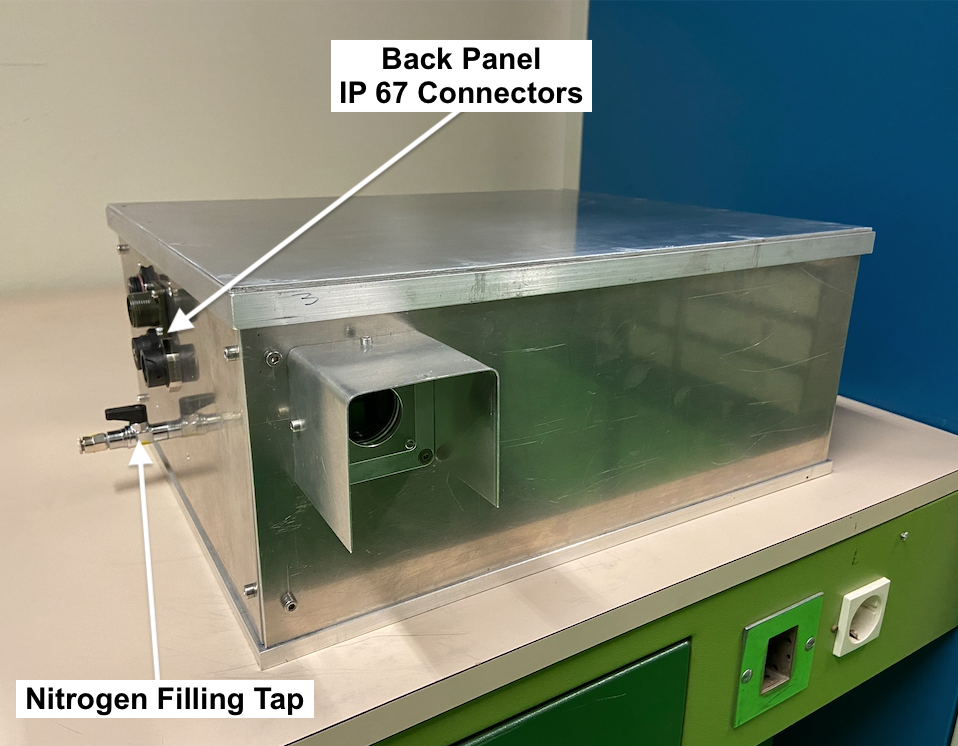}
\caption{External view of the CaliBox. The main frame is protected by an aluminum shell structure designed to be  IP67 waterproof. A brim protects the borosilicate window against rain drops and dust. On the left side is visible a panel to connect the power supply, the Ethernet cable, the optical fiber linking the trigger board to the Camera and the gas tap for the Nitrogen.}  
\label{fig:external_view}
\end{figure}

Outside the two boxes there are an ODROID-C1+ mini PC, a 100 Watt power supply and the CaliBox trigger interface board. An aluminum weatherproof shell, IP67 certified, protects the electronics components and the two boxes  as shown in Fig.~\ref{fig:external_view}. The shell  is designed so that we can remove its walls to access the electronic components  when the maintenance is necessary. An RH, temperature and pressure sensor is located inside the box containing the laser and another one is inside the shell close to the electronics to provide a continuous monitoring of the environmental conditions, including the dew point.
The electronic components and the laser are in direct contact with the aluminum frame to improve  heat dissipation. To operate in a high external RH environment, a perimeter gasket and a thin layer of marine silicon seal the internal edges of the box. 
\subsection{Optical system}
The optical system is composed of a 1 $\mu$J UV Q-switching  laser emitting a 355 nm beam with a  pulse width of 400 ps (FWHM)  and manufactured by Teem-Photonics  \cite{laser}. The laser is driven by a DC voltage controller board, MLC-03A-BR1, connected via a parallel socket to set the laser pulse frequency in a range of 1-2000\,Hz and to provide a TTL external trigger. As shown in Fig. \ref{fig:optical_path0}, one box contains, in addition to the laser and the controller, two Thorlabs FW102C  filter wheels  \cite{wheel}, WH, placed in series. Each is equipped with two sets of five different UV Neutral Density (ND) filters, leaving one filter window in each wheel empty. The attenuation percentages of the two sets of filters mounted in WH1 and WH2 are 39., 11.0, 3.0, 0.5, 0.053 and 31., 15., 7.8, 1.6, 0.034 \% respectively. 
The 35 filter combinations provide a  photon flux  from a single photon up to $10^5$ photons sent to the Camera PMTs. 
The filter wheels are inclined by a few degrees with respect to the wall of the CaliBox in order to avoid  reflection of the laser beam by the filters going back into the laser optical cavity. The light pulse generated by the laser,  passing through the two filter wheels, enters the 1-inch Ulbricht sphere located inside the  separate aluminum box as described in Sec. 2.3. The sphere diffuses  the laser light uniformly, thanks to the inner  thin layer of Spectralon, over an opening angle of $\sim$ 6 degrees. An aluminum diaphragm of 15\, mm diameter collimates the light beam on the LST-1 Camera 28 m away over an area of about 4\, m$^{2}$. A sealed borosilicate  50\, mm diameter window by Edmund Optics  allows  98\,$\%$ of the UV light to be transmitted to the Camera. During the data taking, a filter combination that provides about 80 p.e./PMT is used to satisfy both high and low PMT gain readout, avoiding the saturation of readout electronics \cite{phec}.

\begin{figure*}
\centering
\includegraphics[width=.7\textwidth]{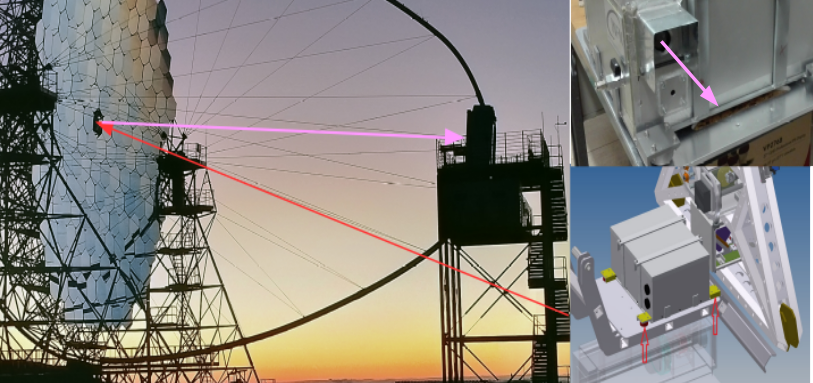}
\caption{Left: The LST-1 telescope.  Right:   The CaliBox (top) and (bottom) the design to show how the CaliBox is installed at the center of the telescope dish. The red arrows show the vibro-stops installed to reduce the vibration of the CaliBox frame.  }  
\label{fig:optical_path}
\end{figure*} 

\subsection{Monitoring of the photon flux}
Inside the box containing the Ulbricht sphere, a system is installed to monitor the photon flux sent to the Camera. 
It is composed of a beam splitter,  two proximity photosensors and a 355 nm band pass filter, BP to filter the external light from the environment, as shown in Figs. 4. The diffuse light coming out of one of the two exits of the sphere is sent to two sensors using a beam splitter, one for low photon intensity, SenSL 3x3 mm$^2$ Silicon-Photomultiplier,
SiPM, and  for high photon intensity, a photodiode (PD) as shown in Fig. 4. For a filter combination corresponding to an optical density less than $10^{-3} $  the SiPM sensor detects the light pulse while for higher intensity it is switched off. According to the filter combination used, the analog signal detected by one of the two sensors is read by a 12-bit ADC allowing to evaluate the number of photons sent to the Camera as discussed in Sec. 5. The monitoring of the photon flux  is implemented in the OPC-UA protocol to periodically provide  information  and to verify the possible degradation of the internal optical components over time.

\subsection{Computer Board and Electronics}
The CaliBox is controlled by a ODROID-C1+ equipped with Ubuntu 14.04 LTS Linux system as shown in Fig. 1. This choice is due to its high versatility for the presence of four (plus one) independent Universal Serial Bus (USB) ports and a General Purpose Input Output (GPIO) interface to allow the analog-digital communication with all the internal devices. 
The USB protocol, as shown in the schematic of Fig. 1, is used to control two USB relays placed outside the optical frame: the laser relay to protect the laser controller power up and  the VR relay to control the SIPM power supply. The other three USB ports are used to drive the two filter wheels, to select the desired filter combination and to arm and disarm the laser. 
All the subsystems connected by USB ports include hardware feedback to know their status and provide all data in real time to the OPC-UA system. 
A dedicated  GPIO port triggers via TTL signal  the laser with a user defined pulse repetition frequency,  duty cycle and  number of cycles.  The temperature and relative humidity of the two ambient sensors (RH/P/T) inside the CaliBox are also read by the Serial Periferal Interface (SPI)  bus. 
During the data taking, we use the TTL trigger provided by the laser controller to synchronize the pulse with the Camera electronics. This TTL signal is converted to an infrared  optical signal using a Vertical Cavity Surface Emitting Laser (VCSEL) and sent over optical fiber to the trigger interface board of the camera (TIB) and  translated into a TTL standard signal  by a dedicated photosensor. This information is then used to trigger the readout of all camera pixels \cite{tibcam}.
To get the PMT charge distribution without the laser pulse, a random TTL signal generated by the ODROID-C1+ is  sent to the  TIB  to  evaluate the PMT pedestal.
To monitor the flux of photons coming out of the Ulbricht sphere, described in Sec. 2.3, a 12-bit ADC is connected by SPI to four GPIO dedicated pins and triggered by the laser TTL signal using an Interrupt Request (IRQ) sent by the ODROID C1+.
An interlock relay  is activated when the entrance door of  the telescope area is open and it inhibits  the laser for safety. The OPC-UA server includes the possibility to set several Methods (MeTs) and hardware-software feedback by DataPoints (DP) to monitor all the parameters set.

\begin{figure*}
\centering
\includegraphics[width=.7\textwidth]{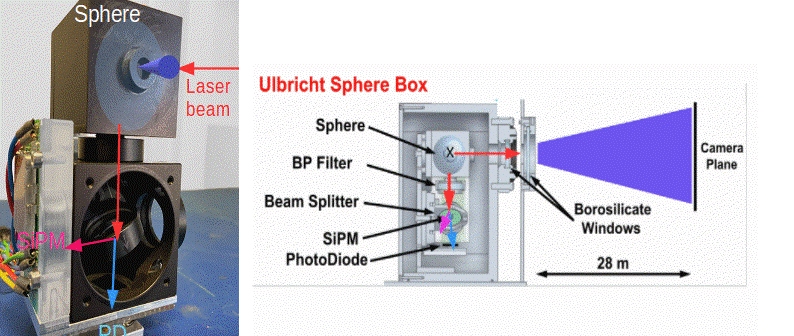}
\caption{Optical internal calibration structure. Left: The cube on the top houses the diffuse sphere. The cube on the bottom bears the beam splitter. The figure shows the SiPM which collects photons reflected by the beamsplitter and the PD (at the bottom of the beamsplitter cube) which monitors the transmitted photons.
Right: The schematic description of where the components of the optical internal calibration are located. }  
\label{fig:optical_path}
\end{figure*} 

\section{Pulse shape and estimation of the photon flux variability}
\begin{figure}
\centering
\includegraphics[width=.9\columnwidth]{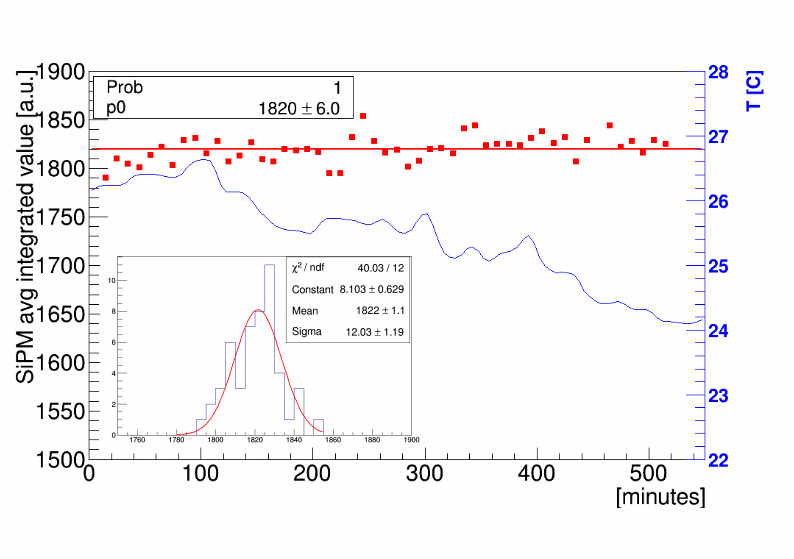}
\caption{Scatter plot of the laser beam average signal amplitude  versus time measured  every 10 minutes over nine hours (red squares). During this test,  the laser pulse was set  to 100\ Hz. The linear fit  shows laser light stability below 1.0\,$\%$. The variation of the temperature in the Laboratory is also shown, blue line. The insert shows the Gaussian fit of the averaged integral values.}  
\label{fig:ls1kHz}
\end{figure}

To calibrate the Camera we need a signal with the similar characteristics as the pulse generated by the Cherenkov light produced by the air shower in a  single PMT. The FWHM of the Cherenkov signal is about a few ns. The width of the pulse generated by our system depends at the first order on the diameter  and reflectivity of the diffuse sphere. To get a reasonable FWHM compared to the Cherenkov signal, we hence have used a 1-inch Ulbricht sphere. By using a PMT, which has similar performance as the camera ones, located in front of the CaliBox window exit at 5 cm from the center of the Ulbricht sphere,  we have measured the laser pulse FWHM to be $2.80\pm 0.03$ ns, consistent with the Cherenkov LST camera signal \cite{icrc2019}.
Another relevant parameter of the calibration system is the stability of the photon flux sent to the camera.
We have measured the stability of the beam intensity emission  every 10 minutes over nine  hours using a SiPM sensor installed on one side of the splitter and a PSI-DRS4 evaluation board signal digitizer \cite{drs4} to compute the signal pulse integral per event. The gain of the SiPM was adjusted as a function of the temperature read by a sensor located inside the diffuser box connected to the CAEN voltage controller.  A sample of  1000 events was taken using a 100 Hz laser pulse frequency for nine hours to reproduce the same data taking a condition of a night shift. Fig.\,~\ref{fig:ls1kHz} shows the average integrated pulse signal versus time is stable  within $1.0\pm 0.2 \%$ (rms over mean). The peak-to-peak stability defined as the percentage  difference between  the minimum and maximum  values of output power  and  the average power, representing the range of output power variation within nine hours, results to be equal to 1.2 \%. This factor takes into account the output power fluctuation related to the  temperature drift, pump power fluctuation and the crystal temperature change. In the measurement done in Laboratory the ambient temperature  plotted in Fig. 5 shows  there is no relation between the temperature variation in the Laboratory  and the beam laser intensity thanks to the fact that the optical system is inside a hermetically sealed box. 

The PMT camera calibration requires a high photon dynamic range because the number of the p.e. per pixel is related to the energy of the primary, its type (hadron or gamma) and its impact parameter.
Hence, we have performed in Laboratory a test to evaluate the photon density variability reaching the camera plane as a function of the filter combinations. The measurements were done using filters with optical density ranging from 1 to 3 and the signal was detected by a SiPM. 
The sensor is mounted on one side of the splitter, as shown in Fig. 4, and its gain, primarily influenced by temperature, was stabilized over time. The bias voltage was continuously adjusted by a CAEN voltage regulator board, with real-time updates provided by a microprocessor using the temperature value measured by a temperature sensor installed inside the diffuser box.
 The signal was recordered and  digitized by the PSI DRS4 evaluation board.  The method used to evaluate the photon density is based on the assumption that the number of  photons per pulse  released by the laser is given by a Poisson distribution and the probability to be detected by the  sensor  is described by a binomial distribution. Under these assumptions the number of detected photons by the sensor is described by  a Poisson distribution where the pulse integral variance  $\sigma^{2}$, for different optical densities is proportional  to the number of p.e.. 
 Fig. 6 shows the dependence of $\sigma^{2}$ on the integrated signal mean  for different filter optical densities and the slope of the linear fit is equal to   131.1 $\pm$ 2.4 \,Vns/p.e.
Taking into account the filter density combination  to reduce the light intensity  sent to the sensor, a PDE of 20$\%$ of the sensor at 355 nm, a splitter of laser  light of 30$\%$ and the effective sensor surface  we  evaluate the number of ph/cm$^{2}$ at the distance of  6\,cm from the center of the Ulbricht sphere to be  $(3.3 \pm 0.2) \times 10^{7}$ ph/cm${^2}$.  This value guarantees the dynamic range of photons sent to the camera as requested by the technical project, \cite{doc1}.

\begin{figure}
\centering
\includegraphics[width=.8\columnwidth]{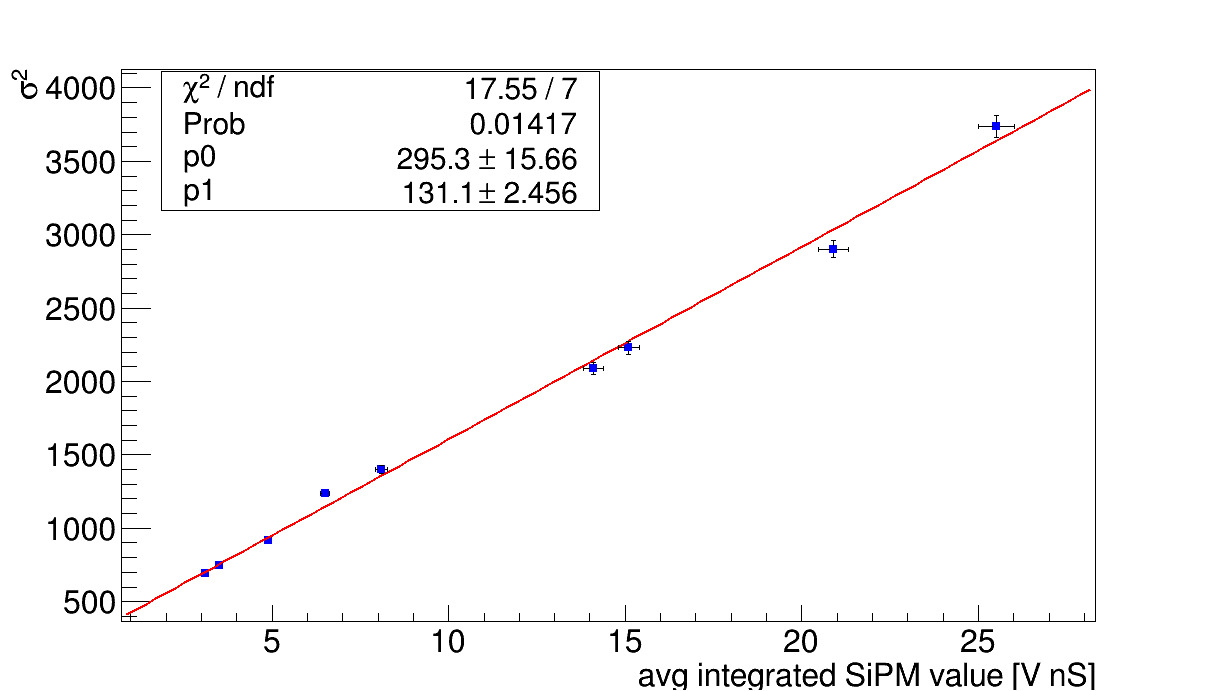}
\caption{Signal integral variance, $\sigma^{2}$, versus mean SiPM  integrated signal measured with several filter combinations ranging  between 3-4 filter optical density. The parameter, p1, of the linear fit  is equal to the pulse-integral/pe.}  
\label{fig:kfactor}
\end{figure} 

\section{Photon uniformity measurement}

 For the energy shower reconstruction, a high precision flatfielding is required to equalize all the PMTs response  and the pattern of illumination across the full camera field of view must be known at the level of 2\% during the data taking. This requirement can be achieved with good optical alignment with the camera and good stability of the photon flux emitted by the laser.

\begin{figure}
        \centering
 \includegraphics[width=.9\columnwidth]{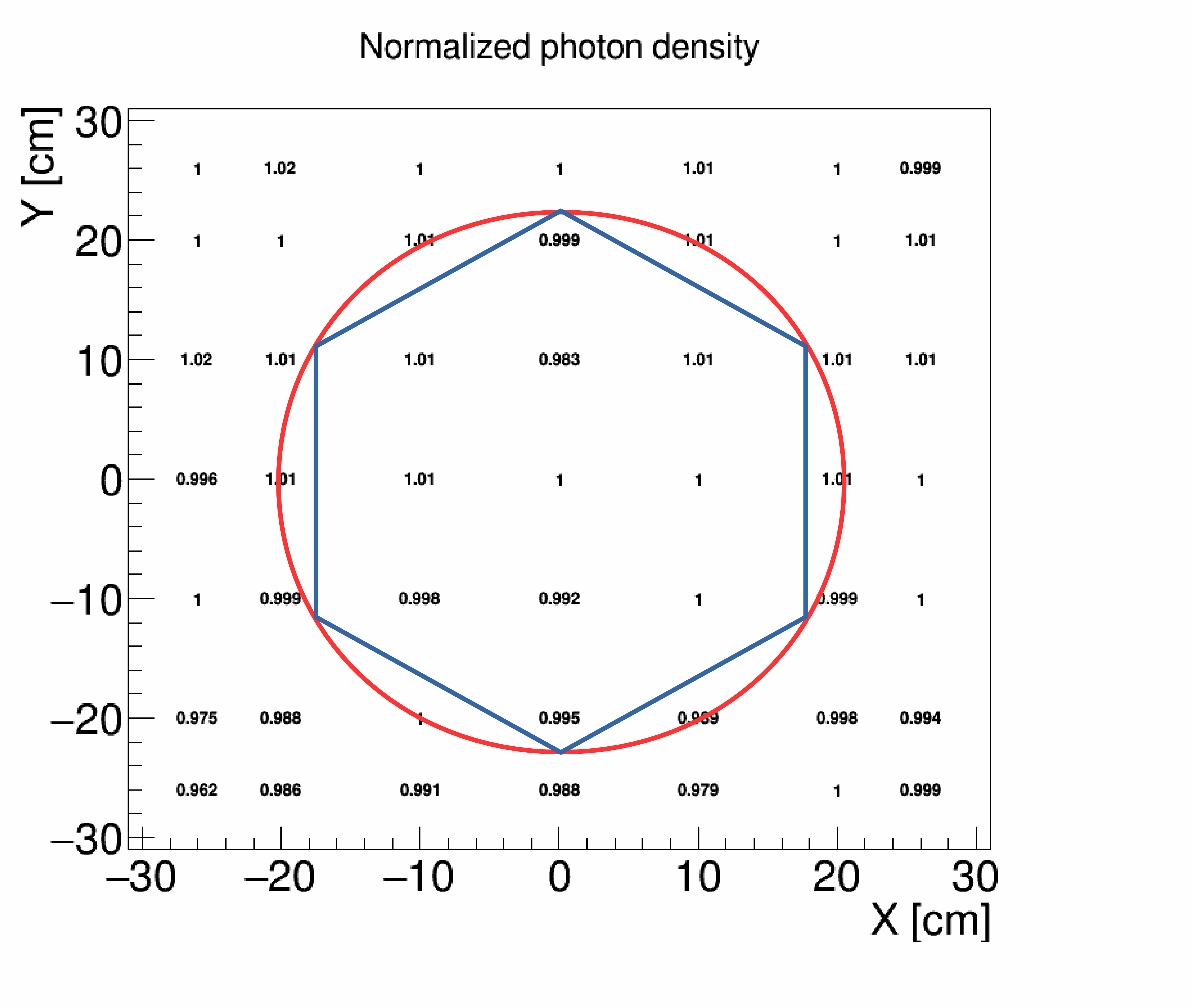}
    \hfill 
    \label{fig:3}
    \caption{Normalized photon density for different $(x,y)$ points. The values, normalized to $(0,0)$,  range between 0.98 to 1.02 with an average value of 0.999 $\pm$ 0.001. }
\end{figure}  

For this purpose we have measured in Laboratory the CaliBox uniformity  using a solid-state 3$\times$3\, mm$^{2}$ SiPM sensor with high detection efficiency at 355\,nm to collect the laser pulses. A rigid panel has been placed 5\,m away from the CaliBox exit window and shifted on a $xy$ plane allowing us to estimate the uniformity at 5 m  within a field of view of about 3.4 degrees capable of illuminating the camera plane 28 m away. The sensor has been  placed at different positions in a  Cartesian  $(x,y)$ coordinates as shown in  Fig.\,7, spanning a distance of $\pm$ 27\,cm  along the $x$ and $y$ axis. For each $(x,y)$ position, 1000 events have been taken, each event corresponding to a laser pulse detected by a SiPM. Every signal was digitized and recorded on a hard drive using a PSI DRS4 evaluation board managed by a dedicated software. Before the data taking, the CaliBox was aligned  respect to the panel by a laser pointer. After the acquisition we evaluated and computed the integrated SiPM signal $\rm{I_{s}}$ for each $(x,y)$ coordinate position computing the average $\langle\,\rm{I_{s}}\rangle$ over 1000 events.
\begin{figure}
        \centering
        \includegraphics[width=.9\columnwidth]{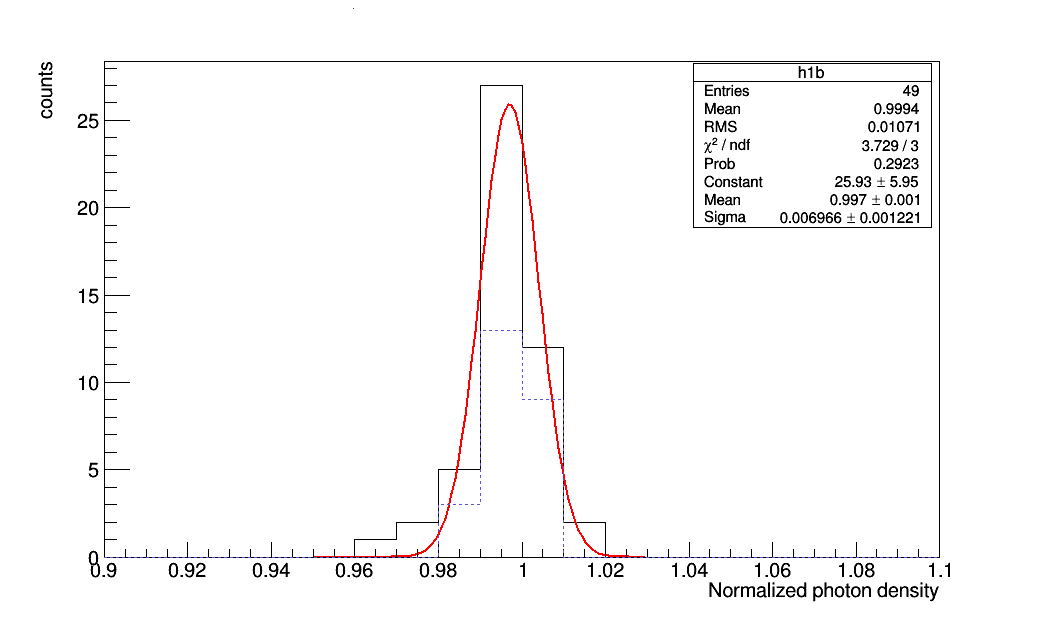}
    \label{fig:3}
    \caption{ Normalized dispersion of mean charge read by a sensor
    on the plane 5 m from the CaliBox as shown in Fig 7. The dotted line shows the normalized dispersion of the mean charge value inside the circle drawn in Fig. 7. The $\sigma$ value gives at first order the variation of photon density.}
\end{figure}  

The evaluation of the uniformity of the illumination, at first order,  is given by the dispersion of  $\langle\,I_{s}\rangle$ over the panel estimated by fitting a function to the distribution. Fig. 8 shows the data are well fitted by a Gauss function, allowing us to estimate the uniformity by $\frac{\sigma}{\langle\,I_{s}\rangle}$,  to be 0.70 $\pm$ 0.10 \,$\%$. 
This value provides the high quality of the optical elements and their precise alignment  and however guarantees that we satisfy the technical requirements of the project to estimate.

\section{Photon flux monitor}
A purpose of the  CaliBox, as discussed in section 2.3, is to monitor the flux of the photons sent to the camera during the data taking by two sensors, SiPM and PD.
For the lower flux, with a filter combination corresponding to a transmittance lower than 10$^{-3}$, only the SiPM is sensitive enough for the monitoring, while for the higher flux, the PD is preferred.
The signal detected by the sensor  with specific wheel combinations is integrated by a 12-bit ADC, triggered by the laser signal. Due to the SiPM signal duration of 50 ns  we have designed and built at INFN Rome1 Electronics Laboratory, LABE, a peak-hold circuit and the output of an operational amplifier, OPA,  is shaped  to an exponential signal with an amplitude defined by the peak of the signal and $\tau$  about 200 $\mu$s.
A TTL signal generated by the ODROID pulses the laser and  generates an IRQ signal  to set the ADC chip select, CS. The ADC value is read via SPI. The timing delay between trigger signal  and IRQ  is adjusted by software. The precision of the ADC measurement is defined by the  CS time position  in  the exponential signal generated by the peak hold circuit. The  measured average jitter of CS  is  about 100 $\mu$s.  As the CS jitter distribution is flat in time, the variance related to the statistics of the incident photons still follows the Poisson statistics.  Hence, by stepping through several filter combinations, we expect a linear dependence between the variance and the ADC counts as suggested by a Poisson distribution.

Fig.\,9  shows, respectively,  the linear relation between the peak height of the SiPM  signal  measured and the ADC counts where the pedestal is subtracted.
The linear dependence between ${\sigma^{2}}_{ADC}$   versus the ADC counts, shown in Fig. 10, allows to evaluate the number of photoelectrons per ADC channel. By using the relationship between the ${\sigma^{2}}_{SiPM}$, of Fig. 6, and the average value of the SiPM pulse integral, we calculate a conversion factor in photoelectron per ADC channel of $4.2\times 10^{-4}$ p.e./ADC.

\begin{figure}
        \centering
        \includegraphics[width=.9\columnwidth]{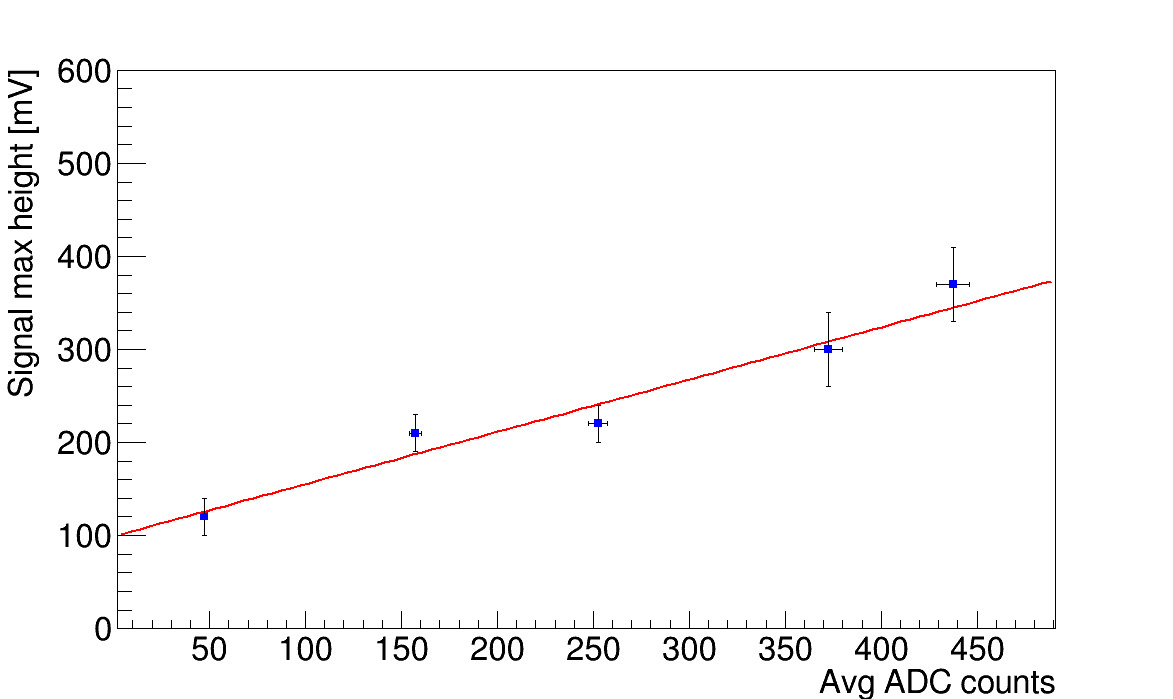}
    \label{fig:3}
    \caption{  The maximum amplitude of a SiPM  signal input to the peak holder circuit versus the ADC counts after pedestal subtraction.}
\end{figure}  

\begin{figure}
        \centering
        \includegraphics[width=.9\columnwidth]{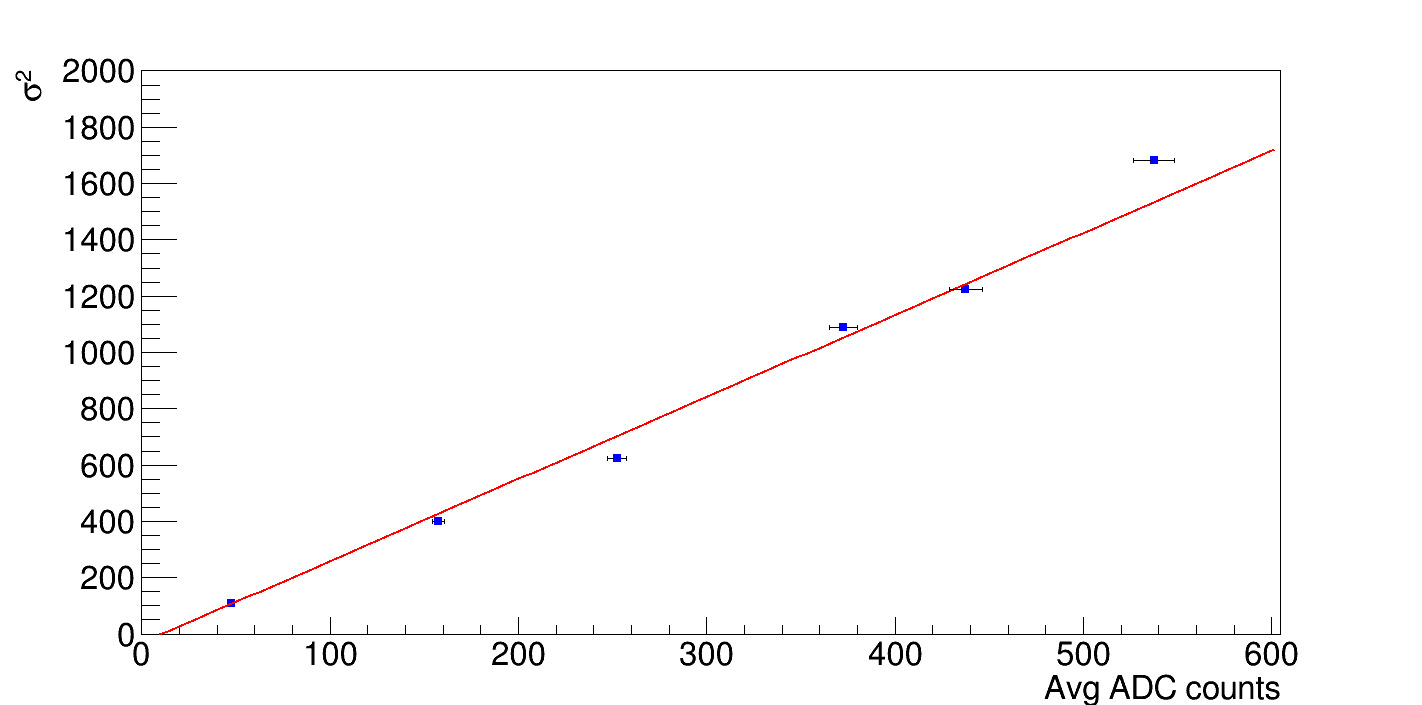}
    \label{fig:new3}
    \caption{  The $\sigma^{2}$ of the ADC measurement versus the  ADC mean measured with several filter combinations ranging  between 3-4 filter optical density. The error includes the error due to the jitter of the CS. The conversion factor in photoelectrons per ADC channel is $4.2\times 10^{-4}$. p.e/ADC.  }
\end{figure}

\section{Test results on thermal conductivity and relative humidity}
Several tests were performed to verify the behaviour of the CaliBox in presence of the extreme environment condition, namely RH as high as 96 \% and temperature ranging from -10 to 30$^{\circ}$C.
 For this purpose, we have used a climatic chamber and sensors  to record temperature and  relative humidity outside and inside the CaliBox. The CaliBox filled with Nitrogen at atmospheric pressure was inserted  into a climatic chamber. The laser was pulsed for two hours while data acquisition was running. An open bowl filled with water was placed in the climate chamber to produce a relative humidity of 96\%. Fig. 11 shows the variation of relative humidity and  temperature as  a function of  time inside the CaliBox when the climatic temperature was set to 30$^{0}$C.  The relative humidity inside the CaliBox stays constant and slightly drops when its internal temperature increases  because the laser is firing.

\begin{figure}
\centering
\includegraphics[width=.9\columnwidth]{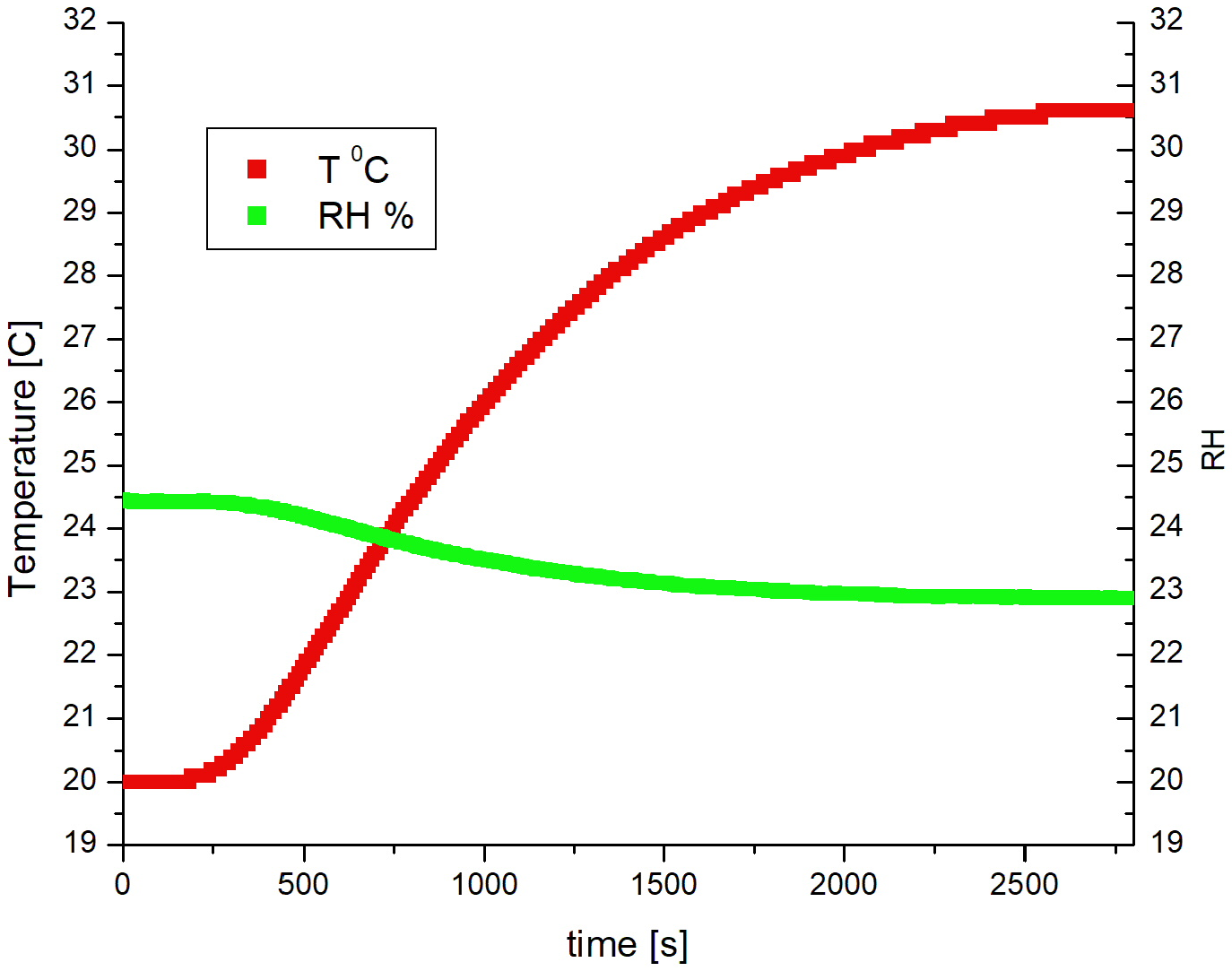}
\caption{Temperature, (left Celsius degrees, red) and relative humidity (right scale, green), respectively  measured inside the box containing the optical system and filled with Nitrogen as function of time.The RH ranges between 24.5\% and 23.0\%.The humidity inside the climatic chamber was about 96$\%$ and its temperature  was 30$^{0}$C during the test. The decrement of  humidity is due to the rise of the temperature of the electronics components being the laser firing.}  
\label{fig:humidity}
\end{figure}

\section{Conclusions}

A optical calibration system for calibrating large telescopes has been designed to ensure stability in photon flux measurements by controlling the parameters that can alter it. The results from the Laboratory tests show the calibration system satisfies the requirements  of the CTAO project that requires a stability of the photon flux at 2\%, a dynamic range of 100-400 photons per pulse  as well as the uniformity of the illumination of the camera of 2\% or less.
The measurements, performed in the Laboratory, show the photon flux at the exit of the diffuser has been constant within 1.0 \% in a time interval comparable to the night shift. In the tests of the CaliBox we have measured a pulse width at the exit of a one inch Ulbricht sphere to be $2.80\pm 0.03$\,ns consistent with the Cherenkov signal read by the camera. The measurements in Laboratory show  that the illumination on a panel 5 m away from the CaliBox has a dispersion of $0.70 \pm 0.10 \%$ within a field of view of about 3.4 degrees capable of illuminating the camera plane 28 m away. This value expresses mainly the limitation in the procedure of our measurement and  that shows, even at 28 m distance, the illumination uniformity on the camera plane can be kept within the required 2\% according to an allignment of the CaliBox respect to the camera plane.
The tests  of the  CaliBox in a climatic chamber show the device is free of water vapor condensation since the optics system is isolated from the electronics components by confining the optical components to two hermetically sealed boxes. 
The first prototype of this device has been working since 2018 in the LST1 telescope at ORM and three more devices that include the improvements discussed in this paper have been constructed and will be installed in the LST2-4. A spare device is already built and it can replace the prototype in LST-1.

\section*{\centering ACKNOWLEDGMENTS}
We would also like to thank  G. Chiodi, M. Ciaccafava,  M. Iannone, Dr. A. Zullo of INFN-RM1-Engineering Design Office and  A. Mattei INFN-RM1 Mechanical whorkshop.
One of us, M.I,  is grateful  to Dr. Fulvio De Persio who started this project with me and to Dr. Martin Will  who partecipated in the installation of the CaliBox in LST-1. We are grateful to Prof. Barbara De Lotto, Dr. Diego Cauz and Dr. Michele Palatiello  which made possible the photon uniformity measurement  in the DIMI Laboratory at University of Udine.


\bibliography{mybibfileP}

\end{document}